\documentclass {aipproc}

\layoutstyle{6x9}

\usepackage{graphicx}

\usepackage{natbib}

\newcommand{\be}{\begin{equation}}
\newcommand{\ee}{\end{equation}}

\newcommand{\ba}{\begin{eqnarray}}
\newcommand{\ea}{\end{eqnarray}}
\newcommand{\om}{\omega}

\newcommand{\B}{{\bf B}}

\newcommand{\Bf}{{magnetic field}}

\newcommand{\NS}{neutron star}
\newcommand{\ms}{magnetosphere}
\newcommand{\NSs}{{neutron stars}}

\newcommand\cxo {{\it Chandra}}
\newcommand{\tfe}{1E~1048.1--5937}

\newcommand\eg{\textit{e.g.}}

\newcommand\lo{\mathrel{\raise.3ex\hbox{$<$}\mkern-14mu\lower0.6ex\hbox{$\sim$}}}
\newcommand\go{\mathrel{\raise.3ex\hbox{$>$}\mkern-14mu\lower0.6ex\hbox{$\sim$}}}

\begin{document}
\title{Neutron star magnetospheres:  the binary pulsar, Crab and magnetars}
\author{M. Lyutikov}{address={Department of Physics, Purdue University, 525 Northwestern Avenue
West Lafayette, IN
47907}}

\begin{abstract}
A number of disparate observational and theoretical  pieces of evidence indicate that, contrary to the conventional wisdom,  \NSs' closed field lines are populated by  dense, hot plasma and may be responsible for producing some radio and high energy emission.  This conclusion is based on eclipse modeling of the   binary pulsar system PSR J0737-3039A/B (Lyutikov \& Thompson 2005),  a quantitative theory of Crab giant pulses (Lyutikov 2007) and a number of theoretical works related to production of non-thermal spectra in magnetars through resonant scattering. 
In magnetars, dense pair plasma is produced by twisting \Bf\ lines and associated electric fields required to lift the particles from the surface. In long period pulsars,  hot particles  on closed field lines can be efficiently trapped by magnetic mirroring, so that relatively low supply rate, \eg\ due to a drift  from
open field lines,  may result in high density. In short period  pulsars, magnetic mirroring does not work; large densities may still be expected at the magnetic equator near the Y-point.
\end{abstract}

\maketitle

\section{Secret lives of closed field lines}

Our understanding of pulsars is based on the Goldreich-Julian model \citep{GJ}, which postulates that pulsars are endowed with dipolar \Bf;  most magnetic  field lines close back to the star, while those  originating near the magnetic poles  are open to infinity. All the action, like generation of radio and high energy emission,  is assumed to take place on open field lines. Closed lines are assumed to be nearly dead, only  carrying a particle density equal to the  minimal charge density required by the electrodynamic conditions.

A number of  recent   observational and theoretical  developments indicate that (at least in some parts of \ms)  the real particle  density
on closed field lines exceeds this minimal value by a large factor, of the order of $10^4-10^5$ and that these particles can produce radio and high energy emission, as we describe in these proceedings. 

\subsection{Eclipses in the  double pulsar  PSR J0737-3039A/B}

\begin{figure}[!h]
     \includegraphics[width=0.5\linewidth]{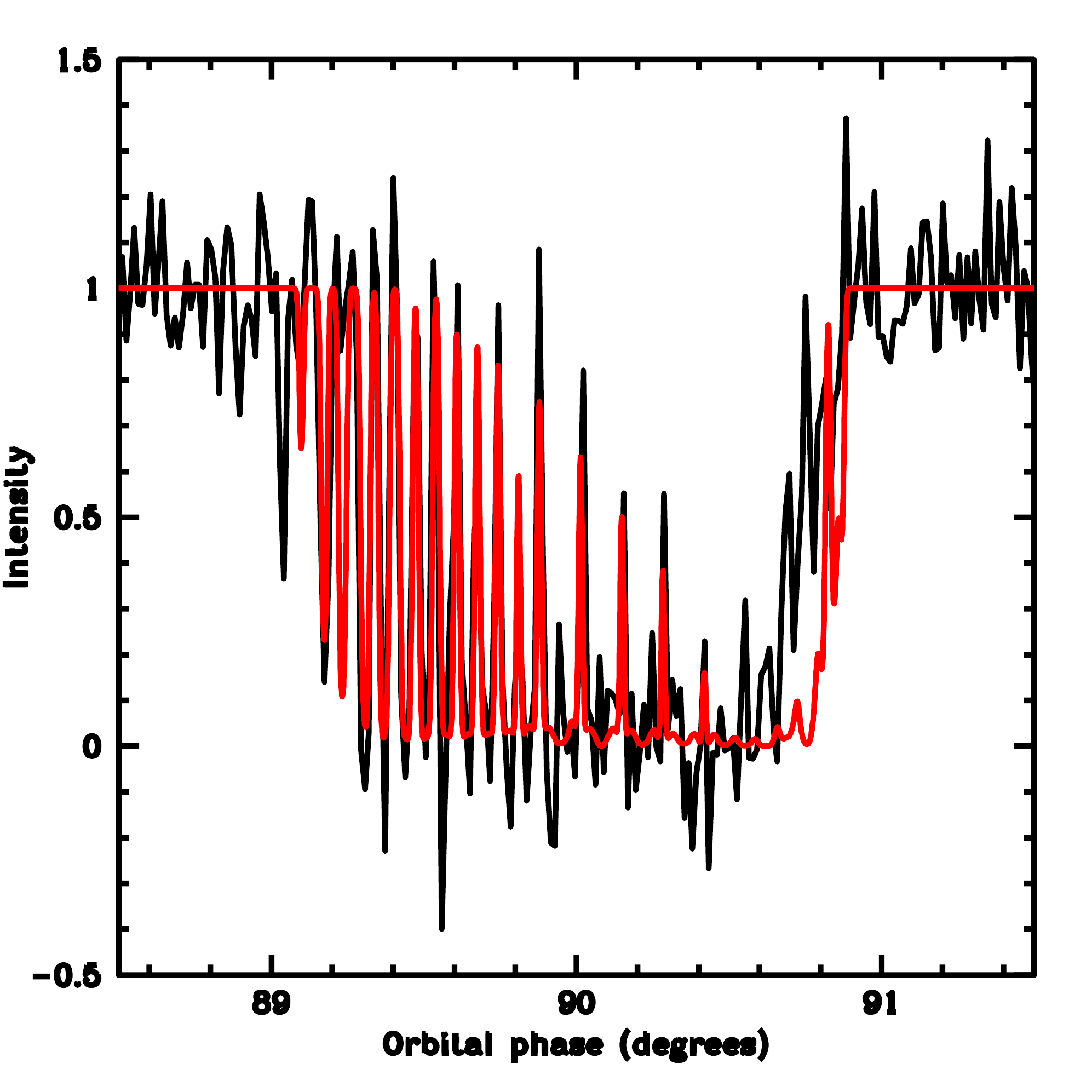}
     \includegraphics[width=0.45\linewidth]{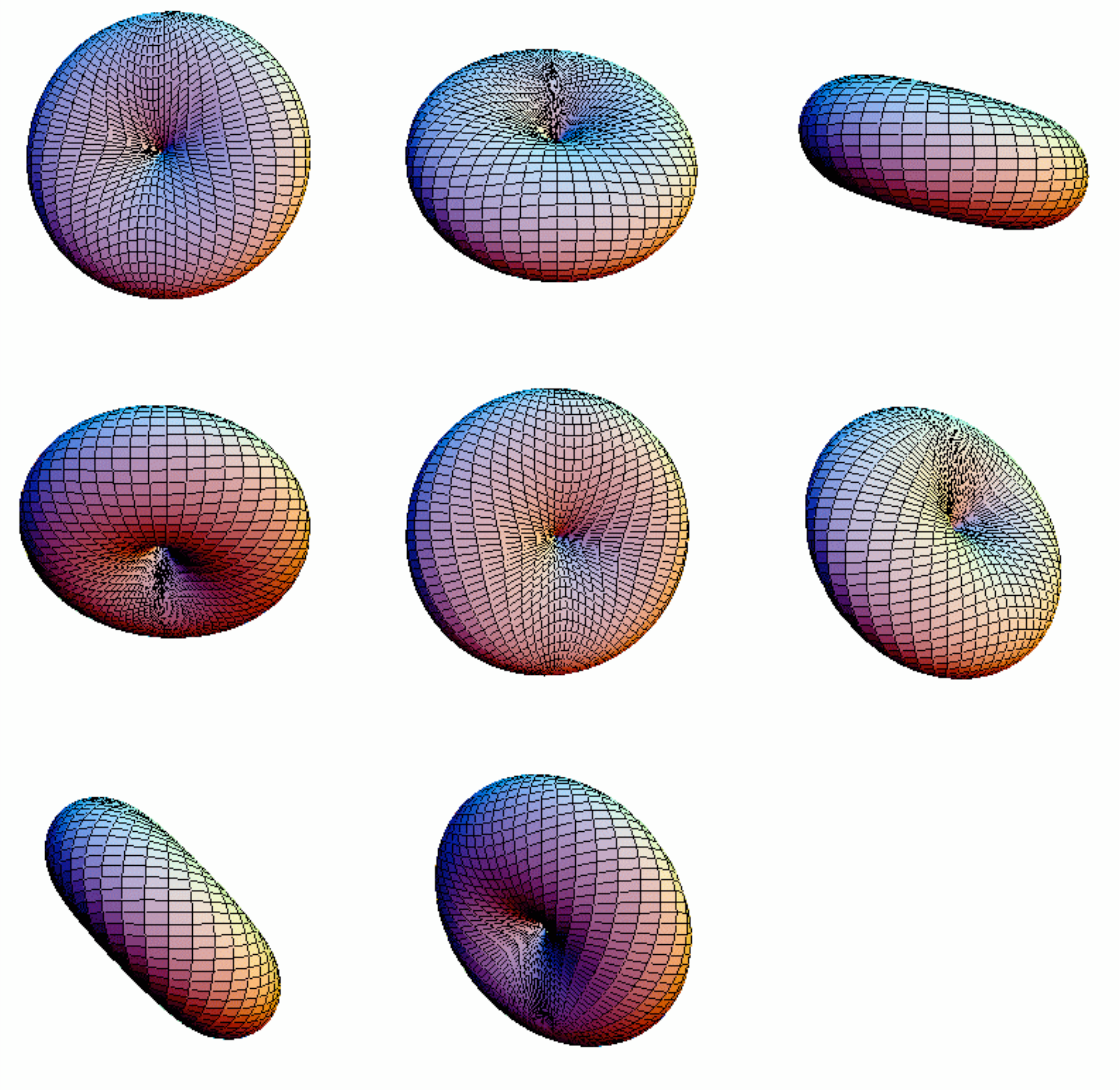}
   \caption{a.
    Comparison of a simulated eclipse profile (red) \citep{lt05} with 800 MHz data (black)
    \citep{mclau04}. 
    The model fits the data best in the middle of the eclipse, where distortions of B \ms\ from the dipolar form  by A wind are small. 
    b. View of the magnetosphere at different rotational phases
 separated by $\pi/4$.  For a full movie of the eclipse see 
 http://www.physics.mcgill.ca/$\sim$lyutikov/movie.gif
 }
 \label{compare}
 \end{figure}

In this system a fast recycled Pulsar A with period $P_A= 22.7$ msec 
orbits
a slower but younger  Pulsar B which has a period $P_B=2.77$ sec in  tightest binary \NS\ orbit
of 2.4 hours \cite{lyne04}. In addition to testing general relativity, this system provides a truly golden opportunity  to verify and advance our models of pulsars magnetospheres,  mechanisms of generation of radio emission and properties of their relativistic winds. This is  made possible by a lucky fact that
 the line of sight lies almost
in the orbital plane, with inclination less than half degree \citep{ransom04,lt05}. Most strikingly, 
Pulsar A is eclipsed once per orbit, 
for a duration of $\sim 30$ s 
centered around superior conjunction (when pulsar B is between the observer and pulsar A).
The width of the eclipse is only a weak function of the observing frequency
\citep{kaspi04}. During the eclipse, the pulsar A radio flux is
modulated by the rotation of pulsar B: 
there are  narrow, transparent windows in  which
the flux from pulsar A rises nearly to the unabsorbed level \citep{mclau04}.
These spikes in the radio flux are tied to the
rotational phase of pulsar B, and provide key constraints on the geometry of the
absorbing plasma. At ingress the
transparent spikes appear at half rotational period of B, then change their
modulation to rotational period of B at the middle of eclipse and disappear completely
right before the egress (Fig \ref{compare}).
The physical width of the region which causes this periodic modulation is
comparable to, or somewhat smaller, than the estimated size of the magnetosphere
of pulsar B.  Combined with the rotational modulation, 
this provides a strong hint that the absorption is occurring
{\it within} the magnetosphere of pulsar B.  This allows us to probe directly 
the structure of pulsar  B magnetosphere.  

The basis for understanding  the behavior of the system is provided by the work 
\cite{lt05},  in which    a
 model  of A eclipses   successfully reproduces the eclipse light curves down to intricate details, Fig. \ref{compare}. The model assumes that radio pulses of A are absorbed by relativistic particles populating B \ms\ through synchrotron absorption. 
 The modulation of the radio flux during the eclipse is due to 
the fact that -- at some rotational phases of pulsar B -- the line 
of sight  only passes through open magnetic field lines where absorption
is assumed to be negligible. 
The model explains most of the
properties of the eclipse:  its asymmetric form, the nearly
frequency-independent duration,
and the modulation of the brightness of pulsar A
at both once and twice the rotation frequency of pulsar B in different parts
of the eclipse. 
 
This detailed agreement confirms the
 dipolar structure of the star's poloidal magnetic field.
One of the most interesting results is that closed field lines are populated with hot dense plasma, exceeding the minimal 
Goldreich-Julian density by a factor $\sim 10^5$.
Presence of  high multiplicity,
relativistically hot plasma on closed field lines of pulsar B is somewhat surprising, but not unreasonable. 
Dense, relativistically hot plasma can be effectively stored
in the outer magnetosphere, where cyclotron cooling is slow.  
The gradual loss of particles inward through the cooling radius, occurring on time
scale of millions of pulsar B periods, can be easily
compensated by a relatively weak  upward flux driven by a fluctuating component
of the current. For example, if suspended material is resupplied at a rate of 
one Goldreich-Julian density per B period and particle residence time is million
periods, equilibrium density will be as high
as $10^6 n_{GJ}$. 
The trapped particles  either drift from open field lines or  heated to relativistic
energies by the damping of magnetospheric turbulence generated by A wind \citep{lt05}.

\subsection{Crab giant pulses}

\begin{figure}[h]
\includegraphics[width=0.5\linewidth] {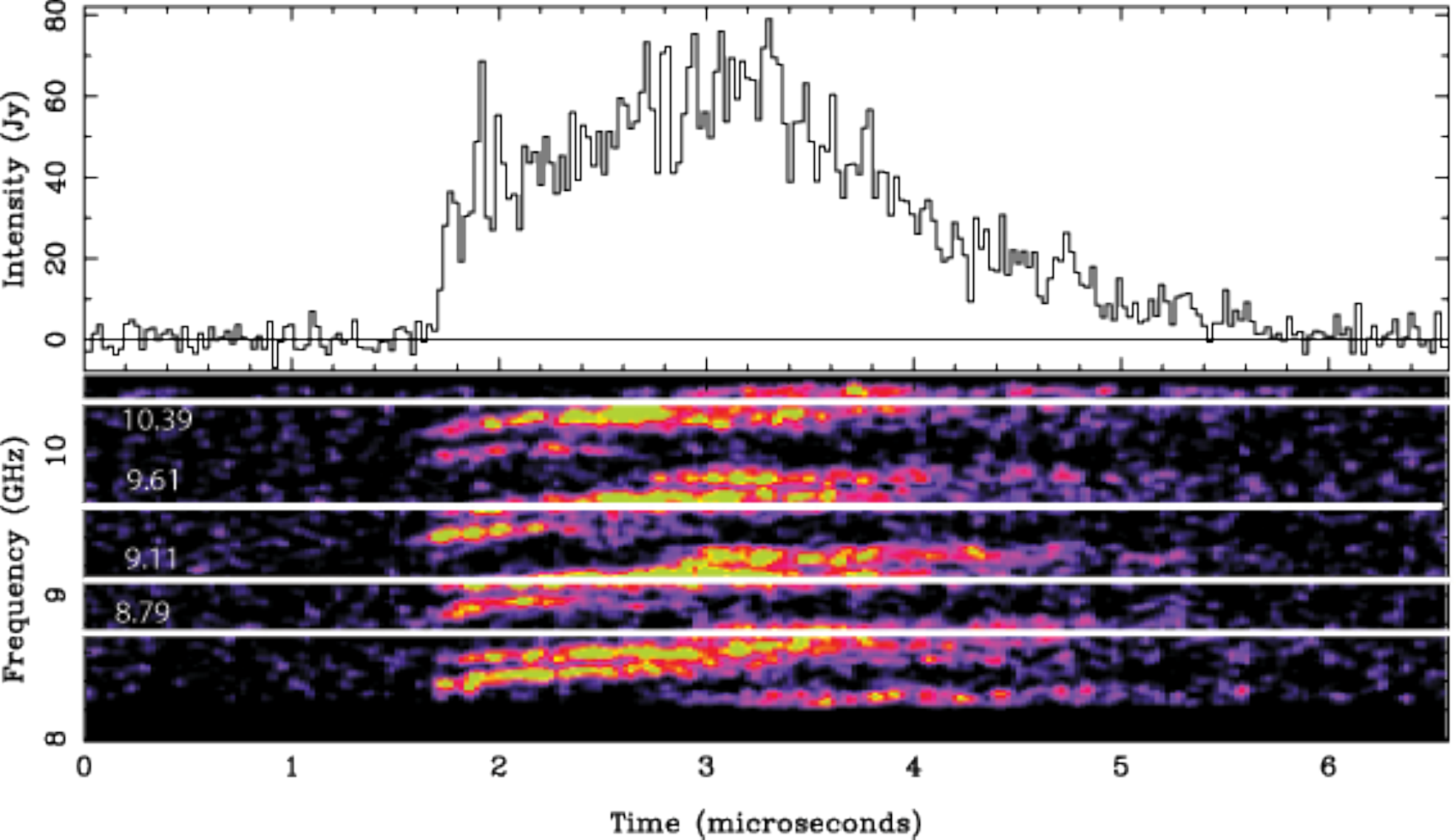}
 \includegraphics[width=.5\linewidth]{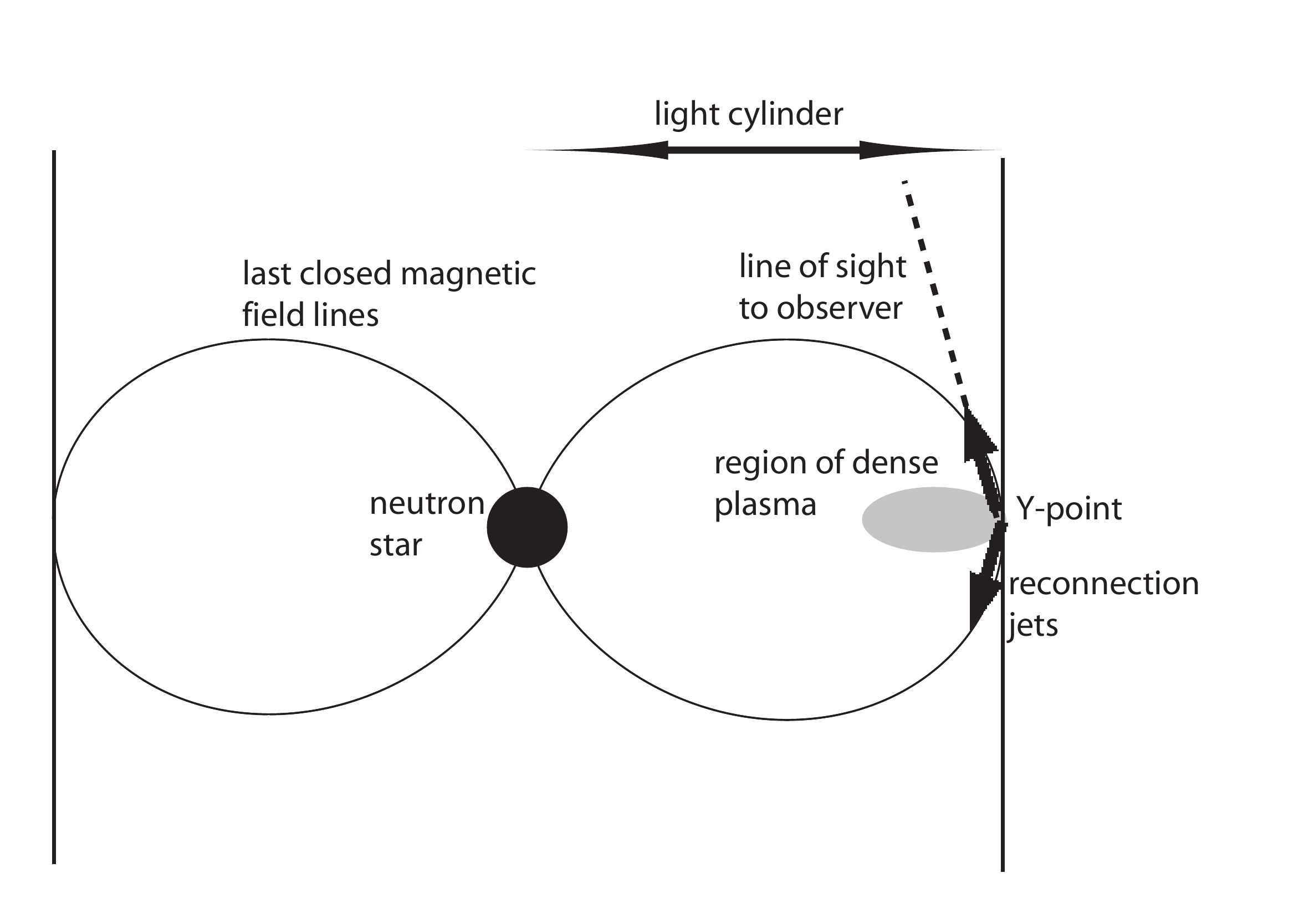}
\caption{ (a) Location of emission bands  (white stripes) for the fiducial model.  The fitted observations correspond to  Fig. 4 of \cite{eilek07}.
(b) Generation region of giant pulses in Crab. High density plasma is trapped on closed field lines near the light cylinder. Occasional reconnection jets produce high Lorentz factor beams that  propagate along \Bf\ lines and  emit  coherent cyclotron-Cherenkov  radiation at  anomalous Doppler resonance.
}
\label{Fit-Bands}
\end {figure}

Another important piece of information about the structure of pulsar  \ms\ and emission processes comes from a series of exceptional high time resolution observation of Crab pulsar at VLA \cite{eilek07}. Over the years the VLA pulsar group observed Crab giant pulses - exceptionally bright bursts of radio emission - with unprecedented temporal resolution reaching sub-nanosecond precision and corresponding large bandwidth. 
This, for the first time in decades,  allows one to study elementary emission processes of  enigmatic  coherent  emission mechanism of pulsars.
To a great surprise,   \citep{eilek07} identified unique features of GPs associated with the interpulse (IGPs)
(Fig \ref{Fit-Bands}): (i) IGPs spectra consist of a number of relatively narrow frequency bands; (ii)  spacing between the bands is proportional, $\Delta \nu/\nu \sim 0.06$; (iii)  emission at different bands start nearly simultaneously, perhaps with a small delay at lower frequencies; (iv)  sometimes  there is a slight drift up in frequencies; all bands drift together, keeping the separation nearly constant; (v)  emission bands are located at $4-10$ GHz, continuing, perhaps, to higher frequencies, but {\it not}  to lower frequencies; (vi) all IGPs show band structure. 

These very specific properties allowed us  to build a {\it quantitative}  model of pulsar radio emission. The model actually does a fitting of pulsar spectra to a particular model \cite{lyut07}, Fig. \ref{Fit-Bands}. 
The model places generation region of  Crab  giant pulses on   closed magnetic  field lines  near   the light cylinder (perhaps, this is  the most striking features of the model,  in stark contrast to all other models of pulsar radio emission). Waves are generated via  anomalous cyclotron resonance  in a set of fine, unequally spaced, narrow emission bands at frequencies much lower than a local cyclotron frequency.  
The model can reproduce  the set of    narrowband, unequally spaced emission bands  seen in GHz frequency range  in Crab giant pulses. 
We stress that this only applies to GPs, regular pulses are generated on open field lines, as  is well established by a multitude of observational facts. We argue that giant pulses are different.
  
  Waves  are emitted at the anomalous cyclotron resonance
\be
\om- k_\parallel v_\parallel = s \om_B/\gamma, \mbox{ for $s < 0$}
\ee 
where $k_\parallel$ and $ v_\parallel $ are components of wave vector and particle's velocity along \Bf,
  $\gamma$ is Lorentz factor of fast particles and
$s$ is an integer. The necessary requirement for the anomalous cyclotron resonance
is that the refractive index of the mode be larger than unity, and that the parallel speed of the particle be larger than the wave's phase speed. The physics of emission is similar to the Cherenkov  process, except that during photon emission a particle {\it increases} its gyrational motion and goes {\it up} in Landau
levels \citep{Ginzburg}. The energy is supplied by parallel motion. Importance of anomalous cyclotron resonance for pulsar radio emission has been discussed in Refs.  \cite{mu79,Kaz91,lbm99}.
 
 A curious  property of anomalous cyclotron resonance is that cyclotron waves can be emitted at frequencies well below the local cyclotron frequency (recall that in Crab near the light cylinder $\om_p/\om_B \sim 2 \times 10^{-3} $). The waves are emitted at the L-O brach (polarized in the $\B-{\bf k}$ plane). Its dispersion depends sensitively on the angle of propagation, qualitatively the resonance condition
 is
  \be
{\om} \sim {  |s| \gamma_{\rm bulk} ^3 \om_B^3 \over \gamma_{\rm beam} \om_p^2} 
\label{X}
\ee
Note, that  for $\gamma  \gg (\om_B/\om_p)^2$ both the  resonant frequency and frequency differences are   much smaller than cyclotron frequency.

    To reproduce the data, it is required that  $r \sim R_{LC}$, $\gamma_{\rm bulk} \sim 1$, implying that plasma is not streaming (closed field lines),
     the  required  density  of plasma on closed field lines is much higher, by a factor $\sim 3 \times 10^5$,  than the minimal Goldreich-Julian density. Emission is generated  by 
a population of highly energetic particles with radiation-limited Lorentz factors $\gamma_{\rm beam} \sim 7 \times 10^7$,  produced  during occasional reconnection close to the Y point, where  the  last closed field lines approach the  light cylinder. The viewing angle with respect to local \Bf is
  $\theta =0.0022$.
 
 The model explains a number of secondary properties of emission bands: (i) that emission bands are not seen below $\sim $ GHz -- the L-O mode  exists only above the plasma frequency; (ii) there is a slight drift up in frequencies, but never down (Eilek \& Hankins, priv. comm) --  we assume that radiating beam is generated close to light cylinder at magnetic equator, where \Bf\ is the lowest; as the beam propagates along \Bf\ lines, the
local \Bf\ increases, leading to increase of resonant frequency, Eq. (\ref{X}). This is similar to frequency drifts of so called auroras hiss in the Earth \ms\ \cite{Labell02}.
 
 The   implications of the model is that 
 closed field lines are populated with dense plasma, exceeding  by a factor $10^5-10^6$ the minimum 
charge density.    Somewhat surprisingly, the over-density $\sim 10^5$ (with respect to Goldreich-Julian density) we find modeling the generation of Crab giant pulsars, is similar to the over-density required to explain eclipses in the double pulsar.

\subsection{Magnetospheres of magnetars}

Magnetars - strongly magnetized \NSs\ - are characterized by powerful persistent and bursting X-ray emission \citep{woods06},
with luminocities well exceeding their spin-down power. Their persistent emission is non-thermal, generally fit with a double black-body or black-body plus power law spectra with a typical photon index of the non-thermal component $\Gamma \sim 2-4$. It was suggested by \cite{tlk02} (see also \cite{BelobThomp}) that large scale currents flowing on closed field lines are driven by twisting of magnetic field lines by the crustal Lorentz stress induced by a tangled \Bf.
These currents results in  particle densities much larger than the Goldreich-Julian
density {\it near the stellar surface}
\be
n \sim \left( {c \over R_{NS} \Omega} \right) n_{GJ} \gg n_{GJ}
\ee
(this estimate assumes that poloidal currents create toroidal \Bf\ of the same order as poloidal \Bf\ and that a typical velocity of charge carries is of the order of the speed of light). 

The presence of such a relatively dense plasma was confirmed by \cite{lg06} through spectral  modeling. The presence of a relatively tenuous plasma leads to large
magnetospheric optical depths for  resonant cyclotron scattering. Due to the large resonant cross-section,
the amount of electron-ion plasma that must be
suspended in the magnetosphere to produce an optical depth of the
order of unity is tiny by astrophysical standards, of the order of
$10^9$ grams total.  Both the emission patterns and
spectral properties of the surface radiation are then modified in the
magnetosphere, as photons are scattered and their energies are
Doppler-shifted in each scatter.

\begin{figure}[h]
 \includegraphics[width=0.55\linewidth]{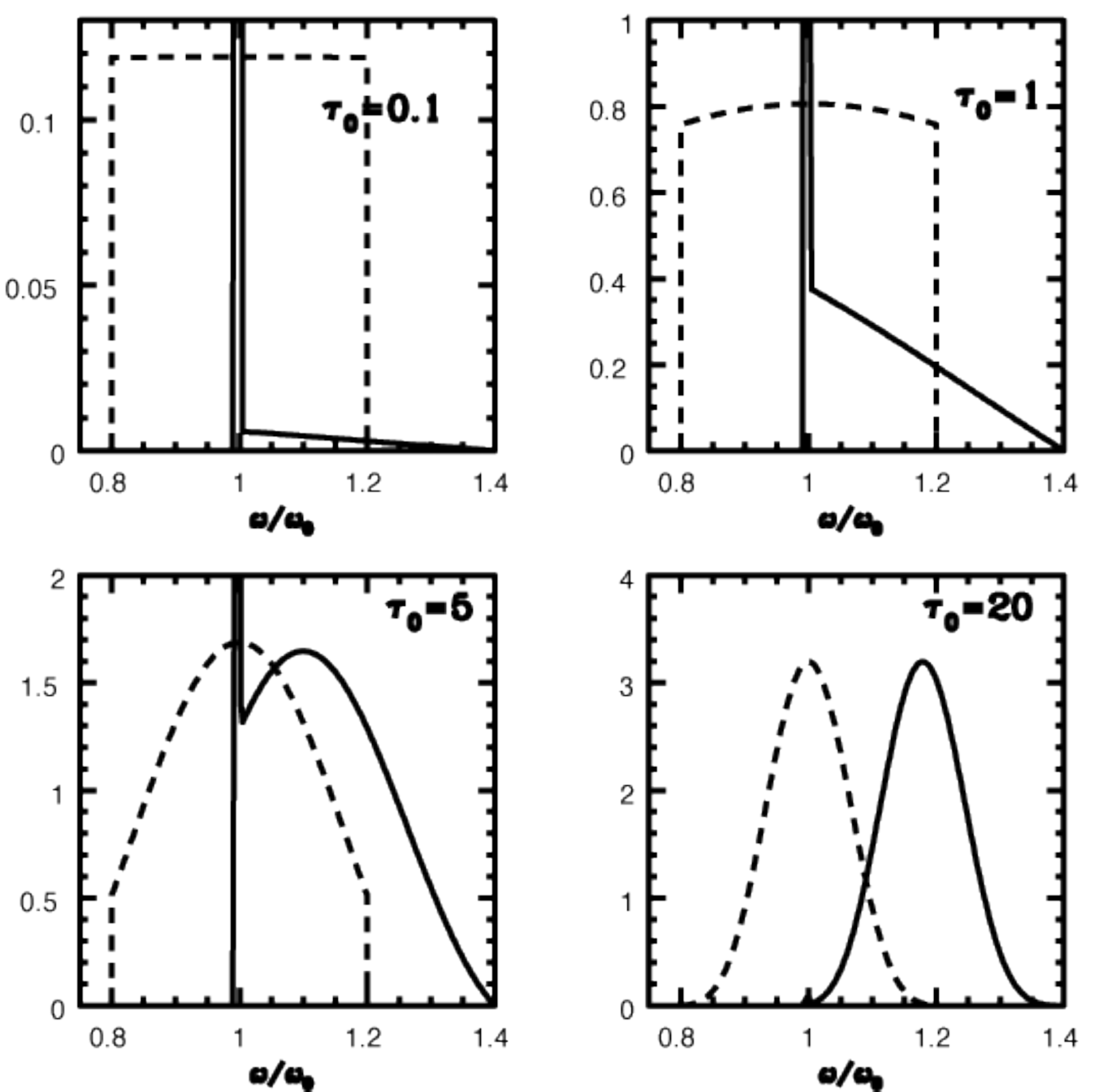}
  \includegraphics[width=0.45\linewidth]{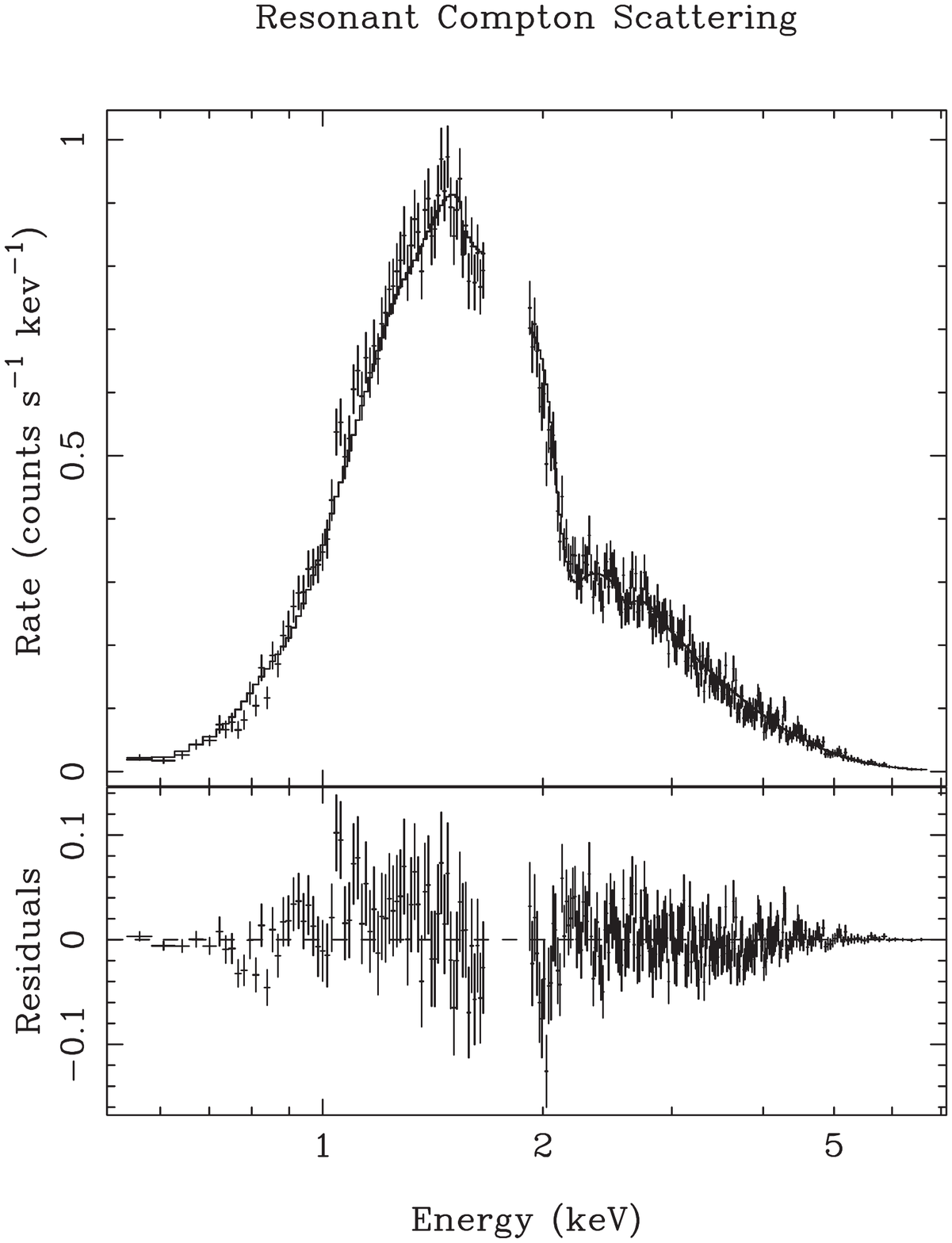}
\caption{(left) Transmitted (solid lines) and reflected (dashed lines) fluxes
for a water-bag distribution of particles with $\beta_T=0.1$ and
different optical depths $\tau_0$. (right) The data fit with the
resonant Compton scattering model to X-ray Spectrum of the Anomalous X-ray Pulsar \tfe\ }
\label{resscat}
\end{figure}

Resonant scattering in {\it  inhomogenous } \Bf\ of \NS\ \ms\ has surprising properties which in a simple Schwarzschild-Schuster (back-forth) approximation and for non-relativistic thermal velocities of plasma particles can be investigated analytically. The scattering of a photon with a given frequency $\om$ occurs in narrow resonant layer centered at
$\om_0 = \om_B$ and a thickness $\sim \beta_T r/3$ ($\beta_T $ is electron thermal velocity in units of the speed of light). If  a surface emission with a given frequency falls on the resonant layer, what are the properties of the transmitted and reflected fluxes? They turn out to be 
  strikingly  different from Thomson scattering and from resonant cyclotron
scattering in a homogeneous \Bf\. (Fig.~\ref{resscat}): (i) in the large optical depth limit $\tau \gg 1$ (so that very few photons pass through the layer without scattering),   the spectral fluxes of
the reflected and the transmitted radiation are equal, so that  {\it a
resonant layer with high optical depth is half opaque}; (ii) the transmitted radiation going from high to low \Bf\ 
is {\it Compton up-scattered by a factor
$\sim 1+ 2 \beta_T$}, while the reflected waves on average have $\om
\sim \om_0$; (iii) as optical depth
increases, the {\it  transmitted and the reflected fluxes have increasingly
{\it narrower}  distributions}, converging back to  $\delta$-functions in the $\tau \rightarrow \infty$ limit.

After calculating semi-analytically the  transmitted flux for initially
Plankian spectra by summing over multiple reflections between resonant
layers, we fitted the result  to \cxo\ X-ray observatory observation of the
anomalous X-ray pulsar (AXP) \tfe.  Our model fairs just as well as the ``canonical'' magnetar spectral
model of a blackbody plus power-law model \cite{lg06}. Most importantly, {\it  the models have the same number of fitting parameters}. 

These results were generally confirmed by \cite{rea07}, who applied the model to the XMM-Newton observations of  \tfe. Applications of the model  to 1RXS J1708-4009 and
SGR 1806-20  are more problematic: these are harder spectra sources, which would require hotter (relativistically hot) plasma, in which case the model becomes inapplicable. In Ref.
\cite{fernandez}   extensive Monte-Carlo simulations of resonant scattering in \NS\ {\ms}s were performed, taking into account  relativistic effects and  polarization, confirming our results.

\section{Discussion}

We have described three very different  pieces of evidence indicating that {\ms}s of \NSs\ are filled with dense plasma. Are these three facts related? Radio pulsars and magnetars are clearly different:
 in case of pulsars we are probing plasma density at the light cylinder, while in case of magnetars - near the surface, regions that have substantially different properties. The supply of dense plasma into {\ms}s of magnetars is probably accomplished by unwinding of crustal \Bf\ and corresponding twisting of magnetospheric \Bf\ \cite{tlk02,BelobThomp}.

Even   radio pulsars  seem to be very different: in slow pulsars, like PSR J0737-3039B, plasma can be efficiently trapped on closed field lines by magnetic bottling (for  $ \sim 10^6$ periods in  PSR J0737-3039B, \cite{lt05}), but not in Crab, where  radiative decay times are too short, $\sim 10^{-4} $ sec at the light cylinder \cite{Luo07}. In Ref. \cite{lt05} it was   proposed that high densities on closed field lines of PSR J0737-3039B may be explained by interaction with the wind of the companion, but similar overdensity in an isolated pulsar questions that possibility. Though there are several ways particles can populate closed field lines
 (\eg\ kinetic drift from open field lines, pair production by the 'backward" beam from outer gaps), the demands  in  case of short pariod pulsars like Crab are pretty high: the mechanism should create an over-density of the order $10^5$ with no efficient bottling. 
 
 In case of Crab,  this over-density is required in a fairly limited region near the light cyllinder, where IGPs are presumably produced. It is somewhat natural to associate this density enhancement with the
 Y-point, where the last closed field line approaches the  magnetic equator. In fact, we indeed may expect high density around that region. 
 First, even in rigidly rotating dipolar \ms,  the charge density diverges at that point \cite{GJ}. 
 Second,
   in case of a more realistic force-free aligned  rotator,  poloidal   \Bf\   also diverges at the Y-point \cite{Gruzinov,Spitkovsky06}. To resolve this divergency, non-ideal effects such as resistivity \cite{Spitkovsky06}  and/or particle inertia  \cite{Komissarov06} should be taken into account.  At the moment, we leave this possibility to further studies.
  

Finally, though formal similarities between  very different physical systems are often superficial and misleading and should be taken with care,  we note that closed field lines of the Earth \ms\ are very active in producing high brightness radio emission
like auroral hiss,  roars and burst \citep{Labell02}.

\bibliographystyle{aipproc}
\bibliography{../PulsarBib}

\end{document}